\documentclass[structabstract]{aa}  
\usepackage{graphicx}
\usepackage{txfonts}
\usepackage{natbib}
\def\farcs{\hbox{$.\!\!^{\prime\prime}$}}
\def\degr{\hbox{$^\circ$}}
\def\arcsec{\hbox{$^{\prime\prime}$}}
\newcommand{\feii}{[\ion{Fe}{ii}]}
\newcommand{\sii}{[\ion{S}{ii}]}
\newcommand{\caii}{[\ion{Ca}{ii}]}
\newcommand{\ioncaii}{\ion{Ca}{ii}}
\newcommand{\nii}{[\ion{N}{ii}]}
\newcommand{\oi}{[\ion{O}{i}]}
\newcommand{\oii}{[\ion{O}{ii}]}
\newcommand{\oiii}{[\ion{O}{iii}]}
\newcommand{\ariii}{[\ion{Ar}{iii}]}
\newcommand{\h}{H$_2$}

\newcommand{\kms}{km\,s$^{-1}$}

\newcommand{\lam}{$\lambda$}

\newcommand{\cmc}{cm$^{-3}$}
\newcommand{\ang}{$\AA$}
\newcommand{\en}{n$_{e}$}
\newcommand{\nh}{n$_{H}$}
\newcommand{\te}{T$_{e}$}
\newcommand{\xe}{x$_{e}$}
\newcommand{\mjet}{$\dot{M}_{\rm jet}$}
\newcommand{\msolyr}{M$_{\odot}$\,yr$^{-1}$} 
\begin{document}
   \title{Physical Structure and Dust Reprocessing \\
in a sample of HH Jets}

   \author{Linda Podio
          \inst{1}
          \and
          Silvia Medves\inst{2}
          \and
          Francesca Bacciotti\inst{3}
          \and
          Jochen Eisl\"offel\inst{4}
          \and
          Tom P. Ray\inst{1}
          }

   \institute{Dublin Institute for Advanced Studies, School of Cosmic Physics,
     31 Fitzwilliam Place, Dublin 2, Ireland\\
              \email{lpodio@cp.dias.ie}
         \and
             Universit\`a di Pisa, Dipartimento di Fisica "Enrico Fermi",
	     Largo Bruno Pontecorvo 3, 56127 Pisa, Italy \\
             \email{silvia.medves@gmail.com}
         \and
             INAF - Osservatorio Astrofisico di Arcetri, Largo Enrico Fermi 5,
             50125 Firenze, Italy\\
              \email{fran@arcetri.astro.it}
         \and
         Th\"uringer Landessternwarte Tautenburg, 
         Sternwarte 5, D-07778 Tautenburg, Germany\\
              \email{jochen@tls-tautenburg.de}
             }

   \date{Received ; accepted }
 
  \abstract
   {Stellar jets are an essential ingredient of the star formation process
   and a wealth of information can be derived from their characteristic 
   emission-line spectra.}
   {We investigate the physical structure and dust 
     reprocessing in the shocks along the beam of a number of classical 
     Herbig-Haro (HH) jets in the Orion and Lupus molecular cloud
     (HH 111, HH 1/2, HH 83, HH 24 M/A/E/C, Sz68). 
     The parameters describing the plasma conditions, as well as the dust
     content, are derived as a function of
     the distance from the source and, for HH 111, 
     of the gas velocity.}
   {Spectral diagnostic techniques are applied 
     to obtain the jet physical conditions
     (the electron and total density, \en\, and \nh, 
     the ionisation fraction, \xe, and the temperature, \te)
     from the ratios between selected forbidden lines.
   The presence of dust grains  
   is investigated by 
   estimating the gas-phase abundance of calcium with respect 
   to its Solar value.}
   {We find the electron density varies between 
     0.05-4~10$^3$ \cmc, the ionisation fraction \xe\, is 0.01-0.7, the
     temperature ranges between 0.6-3~10$^4$ K, and the hydrogen density
     between 0.01-6~10$^4$ \cmc. Interestingly, in the  
     HH 111 jet, \en, \xe, and \te\, peak in the High Velocity
     Interval (HVI) of the strongest working surfaces, confirming the
     prediction from shocks models.
     Calcium turns out to be depleted with respect to its Solar value, but
     its gas-phase abundance is higher than that estimated in the 
     interstellar medium in Orion.
     The depletion is high (up to 80\%) along the low-excited jets, 
     while low or no depletion 
     is measured in the jets which show higher excitation conditions.
     Moreover, in HH 111 the depletion
     is lower in the HVI of the faster shock.}
   {Our results  confirm the shock structure predicted by models 
     and indicate that the shocks occurring along the jets, and presumably
     those present in the launch zone, are 
     partially destroying the dust grains and that 
      the efficiency of dust reprocessing strongly depend on shock velocity.
     However, the high Ca gas-phase abundance estimated in some of the 
     knots, 
     is not well justified by existing models of dust reprocessing in shocks, 
     and indicates that the dust must have been partially reprocessed in the 
     region where the flow originates.}


   \maketitle

%

\section{Introduction}

Although it is widely accepted that jets play 
an essential role in star formation,
there are still many open questions about how they are 
generated and propagate in the interstellar medium. 
Their optical spectra are characterised by emission lines from 
atomic species which are collisionally excited in the
shock waves generated by the interaction of the ejected material
with the interstellar medium or previously ejected jet material.
These lines contain a mine of information on the jet and the shock 
properties. 
Beyond the jet/shock morphology and kinematics, information on the
gas physical conditions can be derived by developing a grid of shock models
the predictions of which are compared with the observed line ratios 
\citep[e.g., ][]{raga86,hartigan87,hartigan94}; or,
alternatively, by using spectral diagnostics techniques 
which are independent of any assumption
on the heating mechanism \citep[e.g., ][]{bacciotti99}.
The application of these diagnostic techniques 
to various datasets allowed derivation of the jet physical/dynamical
structure, highlighting the stratification in the shock cooling regions
\citep{nisini05,podio06}, the variation of the excitation conditions along
the jet \citep{bacciotti99}, across it \citep{bacciotti00,hartigan07}
 and, in a few cases, with gas velocity \citep{dougados00,coffey08,
garcialopez08}.


An important aspect is also the estimate of the dust content in jets,
a topic that has been poorly investigated to date.
Several theoretical studies have explored the dust content and distribution 
in the interstellar medium.
On one hand both observations and theoretical models show that 
the gas-phase abundances of elements like iron (Fe), magnesium (Mg),
silicon (Si), and calcium (Ca) 
are considerably depleted (by a factor of the order of 10$^2$-10$^4$)
with respect to their Solar abundances because of the formation of 
dust grains \citep{baldwin91,dishoeck93,savage96}.
On the other hand the chemical composition, the size and the content 
of dust grains can evolve as they lose atoms to the gas phase because of 
shocks waves. The high energy gas-grain and grain-grain collisions
occurring behind the shock front, in fact,  may lead to the erosion of the 
grain surfaces (sputtering) and/or to the vaporisation and fragmentation 
of grains 
\citep[][ and references therein]{jones94,jones00,draine03,guillet09}. 
Total dust destruction is expected in high velocity shocks such as 
supernova-generated shock waves ($v_s$ $>$ 200 \kms) but there are very few
observations about dust reprocessing in slower shocks such as the ones
occurring along stellar jets ($v_s$$\sim$10-80 \kms in the jet beam).
A few works have investigated the gas phase abundance of Fe in HH jets 
\citep{beck-winchatz94,beck-winchatz96,mouri00,bohm01,nisini02} and
only recently, the analysis has been extended to other refractory species such 
as Si, Ca, carbon (C), 
nickel (Ni), chromium (Cr), and titanium (Ti) 
\citep{nisini05,nisini07,podio06,garcialopez08}.

To further investigate these aspects, 
in this paper we study the distribution of physical parameters
and, in particular, the gas phase abundance of calcium in a sample
of HH jets (HH 111, HH 1/2, HH 83, HH 24 M/A/C/E, Sz68), 
as a function of the distance from the source, i.e. in 
the different working surfaces along the jet beam, and,
where possible, of the gas velocity. 
As we will show, these kind of estimates 
are very important to understand dust reprocessing in shocks and
are also useful to derive constraints on the 
properties of the jet launching region.

This paper is organized as follows:
in Sect.~\ref{sect:obs} we present the observations and the data
reduction process and in Sect.~\ref{sect:diag}
we briefly recall the principle of the so-called BE diagnostic technique, 
which has been used to derive the gas physical conditions along the jets. 
In Sect.~\ref{sect:phys} we present the physical properties derived 
for HH 111, in two velocity intervals, and for the jet from Sz68,
that is analysed here for the first time.
Similar results derived for the other jets of the sample at higher spatial
sampling than in previous studies are presented in the online material.
In Sect.~\ref{sect:dust} we use the obtained physical parameters 
to estimate the presence of dust grains in the jets 
and we discuss the efficiency of shocks in reprocessing the dust 
comparing our results with the predictions of theroretical
models. 
Finally, in Sect.~\ref{sect:concl} we present a summary of our results.\\

\section{Observations and Data Reduction}
\label{sect:obs}

We observed a sample of HH jets 
(HH 111, HH 1/2, HH 83, HH 24 M/A/C/E, Sz68)
in the optical range.
Our spectra were acquired in January 1998 at the ESO 3.6m-telescope
by using the EFOSC2 spectrograph, the GR9 grism and a 1\farcs5 slit width. 
The slit was aligned parallel to the jet axis with position 
angles: 276.7\degr\, for HH 111, 324.7\degr\, for HH 1/2, 
297.7\degr\, for HH 83, 328.1\degr\, for HH 24 C/E, 219.0\degr\, for HH 24 G, 
and 334.9\degr\, for Sz68.  
The spectra cover the wavelength range
from 5600 to 7335 \ang\, with a spectral resolution $\sim$1300 
(FWHM$_{instr}$$\sim$230~\kms).
The spatial and spectral scales are 0.157$\arcsec$/pixel and 0.872\ang/pixel 
corresponding to $\sim$40~\kms.   
   
The spectral images were flat-fielded, sky-subtracted, wavelength and 
flux calibrated. In addition to the standard data reduction process the 
spectral images of the detected lines (\oi\lam\lam6300,6363, 
\nii\lam\lam6548,6583, \sii\lam\lam6716,6731, \caii\lam\lam7291,7324) 
were velocity-calibrated, resampled to the same velocity scale
for all the lines, and corrected for velocity shifts due to atmospheric 
differential refraction.
During the observations, in fact, the slit was aligned along the jet emission
through pre-imaging in the \sii\lam6731 line.
Because of atmospheric differential refraction, however, the peak of the 
emission in the other lines may be off the slit center inducing a velocity
shift in these lines (see appendix from \citealt{bacciotti02} for a discussion
of the uneven slit illumination effect).
The resampling in velocity and the correction for the above mentioned velocity 
shifts allowed us to accurately align the lines in velocity despite the low
spectral resolution.

As shown in \citet{hartigan07}, the comparison between the slit width 
and the projected width of the target is crucial in spectroscopic observations.
If the slit is narrower than the jet (and the lines are broader than 
the spectral resolution) the detected lines are spectrally 
resolved and the velocity resolution is higher for decreasing slit width.
If, on the contrary, the slit is much wider than the jet (and the velocity
dispersion is lower than the spectral resolution) the observations 
produce emission-line images for each line in the spectrum, but no velocity 
information can be recovered. 
The latter is the so-called ``slitless spectroscopy'' 
(see also \citealt{hartigan04}).

Our observations are seeing-limited and the jet widths are spatially 
unresolved. Thus the critical parameter to be 
compared with the slit width is the 
Full Width Half Maximum (FWHM) of the seeing.
This was varying between $\sim$1\arcsec, when observing HH 1/2, HH 83, 
HH 24 M/A/C/E, and Sz68, and $\sim$1\farcs5 when observing HH 111. 

In the case of HH 111, the seeing was comparable to the slit width and 
the acquired spectra show line profiles which are broader than the spectral
resolution.
Thus the HH 111 jet is partially resolved in velocity, allowing the analysis
of its structure along the jet and in two different velocity intervals, as we 
will explain in more detail in the next section.

In the other jets, on the contrary, the slit is wider than the seeing FWHM.
Thus we obtain images of the jet in the different lines. 
This is evident when comparing spectral images of the detected lines 
with pre-imaging acquired with the \sii\, filter.
To avoid the confusion between the spectral and spatial information 
we integrated the line fluxes over the line ``spectral'' profile, 
suppressing the velocity dimension. 

In the next section we will explain in more detail the analysis performed
in the two different cases.

\section{Application of Spectral Diagnostic}
\label{sect:diag}

Our analysis relies upon the dependence of the ratios between 
optical forbidden lines on the gas physical conditions in the jet, i.e. on  
the electron and total hydrogen density, n$_{e}$ and n$_{H}$, 
the hydrogen ionisation fraction, x$_e$, and the 
 electron temperature, T$_{e}$ \citep{bacciotti95,bacciotti99}.

As a first step, we thus computed for all the jets in our sample the line
ratios used in the diagnostics, i.e. \sii\lam6731/6716, 
\nii\lam\lam6548,6583/\oi\lam\lam6300,6363, and
\oi\lam\lam6300,6363/\sii\lam\lam6716,6731.
For HH 111,  which is partially spectrally resolved,  
we divided the spectral images of the detected 
S$^{+}$, O$^{0}$, and N$^{+}$ lines pixel by pixel, thus obtaining 
position-velocity (PV) diagrams of the line ratios 
(see Fig.~\ref{hh111_ratios}).
For the other jets (HH 1/2, HH 83, HH 24 M/A/C/E, Sz68) 
we integrated over the spectral profile 
obtaining the variation of the ratios only as a function of the distance
from the source.
The distribution of the ratios already gives a qualitative idea of the 
variations of the gas physical conditions along the jet and, for HH 111, 
with the gas velocity.
The ratio between the sulphur doublet lines (\sii\lam6731/6716) 
is a tracer of the gas electron density 
up to the critical density of the \sii\, lines 
(\en$\sim$2.5 10$^4$ \cmc), the \nii/\oi\, 
increases primarily with x$_{e}$, and
the \oi/\sii\, ratio is dependent on both x$_{e}$ and T$_{e}$.

To obtain a more quantitative estimate of the gas physical properties 
we use the so-called BE technique \citep{bacciotti99}, that extracts
diagnostic information from the comparison of observed and calculated ratios.
The method is based on the fact that the involved forbidden lines 
are collisionally excited
and that, in low-excitation conditions and provided that no strong 
sources of ionizing photons are present, the ionization fraction of hydrogen is
tightly related to those of nitrogen and oxygen via charge exchange
processes. 
This allows one to easily retrieve \xe\, (and the total hydrogen density)
and \te\, from the comparison of the observed line ratios. 
As discussed in \citet{bacciotti99} this method assumes that the
lines used in the technique are emitted in a region of similar temperature, 
density and ionization fraction. 
\citet{bacciotti99} and \citet{nisini05} carefully verified this
assumption by using the results from shock models \citep[e.g., ][]{hartigan87}.
They showed that the considered forbidden lines have their peak emission in the
same region behind shock fronts of different shock velocity, and verified that
the obtained parameters are representative of the average value that those
quantities assume behind the shock front in the line emission region.

In this paper we use an improved version of the BE code 
in which the values of the collision strength are updated using the results
in \citet{keenan96} for S$^{+}$, \citet{hudson05} for N$^{+}$, 
and \citet{berrington81} and \citet{mendoza83} for O$^{0}$.
This allows us to obtain a better fit of the collision strength for high 
temperatures (\te\, between 2-5~10$^4$ K) 
and thus to trace correctly the plasma physical conditions
in regions of high excitation, i.e.
where the \nii\, emission is comparable to that in the \sii\, and
\oi\, lines, like 
in some knots of
the HH 83 and the HH 24 jets where we could not obtain results in our
previous work \citep{podio06}.

Note that the diagnostic technique uses
ratios between different species and, thus, the obtained \xe\, and \te\, 
may depend on the chosen set of elemental abundances.
However, as stressed in \citet{podio06}, the {\it relative} variations 
of the parameters (i.e. in different knots and/or velocity
intervals) do not depend on the 
choice of an abundance set (see Fig.~1 of \citealt{podio06}) which, therefore,
can be thought as a model parameter.
On the other hand the most recent determinations by \citet{asplund05} (Solar) 
and \citet{esteban04} (Orion) are in good agreement and give rise to diagnostic
results that differ of at most 15\%.
One of the main goal of this paper is the determination of the
calcium gas-phase abundance with respect to Solar. 
Thus, at difference with \citet{podio06}, the Solar abundances estimated by 
\citet{asplund05} are adopted for consistency in all the diagnostic analysis.
In particular, the values of the temperature and the ionization fraction are
larger when using the abundances from \citet{esteban04}.
The uncertainty over the absolute values is not shown in 
Fig.~\ref{hh111_phys},
\ref{sz68_phys}, \ref{hh1_phys}, \ref{hh83_phys}, \ref{hh24_phys}, because
these are intended to be used to analyse the relative variations 
of the parameters along the jet and in different velocity intervals. 
However, we do consider the uncertainty due to abundances choice
when computing the theoretical \caii/\sii\, ratios to estimate the calcium 
gas-phase abundance (see Sect.~\ref{sect:dust} for more details).

Since the observations are seeing-limited, 
i.e. the angular resolution of the data
is much lower than the spatial sampling, we integrated the line fluxes 
over the seeing FWHM (FWHM$_{seeing}$$\sim$1.2$\arcsec$$\sim$8 pixels) before 
applying the diagnostics.
Thus the de facto spatial sampling of the obtained results in all the 
figures is equal to the seeing.
Moreover, we integrated the line fluxes over the spectral profile for the 
spectrally unresolved jets (HH 1/2, HH 83, HH 24 M/A/C/E, Sz68) and
over two velocity intervals for HH 111:
Low (for $v$$>$-100 \kms) and High (for $v$$<$-100 \kms)
Velocity Interval (LVI and HVI, respectively).

For the HH 111 and the HH 1 jet we corrected the observed fluxes for 
reddening using the values of the visual extinction, A$_{\rm V}$, estimated by
\citet{nisini05} and \citet{podio06}, the standard dereddening procedure from
\citet{draine89} and an interpolation of the extinction law derived by 
\citet{rieke85} for the near-IR bands. 
For the HH 83 and the HH 24 jet we have no estimates of 
the visual extinction because the \feii\lam1.64, 1.32, 1.25 lines were not 
detected in our previous optical/NIR spectra \citep{podio06}.
Thus we could not correct the line fluxes for the reddening before to 
compute the ratios and apply the diagnostics.
On the other hand, \citet{bacciotti99} showed that, since the 
lines used in this diagnostic are very close in wavelength, the errors induced 
by the fact that we are not correcting for extinction are at most about 8-10\% 
for the ionization fraction and about 15\% for the temperature 
as long as the visual extinction is lower than $\sim$3.

Once a set of abundances is assumed, the errors affecting the parameters 
obtained through the BE technique are due to the signal-to-noise ratio of the 
measured fluxes and the uncertainty in the determination of 
the A$_{\rm V}$ which is used to deredden the line fluxes.
These are very low ($<$5\%) in the bright knots of the HH 111, HH 1, and 
HH 83 jets which have strong emission in all the lines, and larger 
(up to 50\%) in the fainter knots of the HH 24 and Sz68 jets.



\section{The Physical Structure of the Jets}
\label{sect:phys}

In this section we briefly describe the new information
gathered by applying spectral diagnostics (i.e. the BE technique) 
to our 3.6m/EFOSC2 spectra of a sample of HH jets.

In Sect.~\ref{sect:hh111} we present the physical parameters (\en, \xe,
\te, \nh) for HH 111,
for which we have been able to highlight for the first time the
stratification of the gas physical conditions as a function of
velocity.

In Sect.~\ref{sect:sz68} we present the parameters obtained for the HH 
jet from Sz68 which has not been analysed before with the BE technique.

The other objects in our sample, already analysed previously 
\citep{bacciotti99,nisini05,podio06}, are here re-analysed with 
the improved version of the code and increased spatial sampling,
for an accurate determination of Ca gas-phase abundance along the jets 
(see Sect.~\ref{sect:dust}).
The physical properties of these jets are shown in the online material,
for reference.

 


\subsection{HH 111: the High and Low Velocity Intervals}
\label{sect:hh111}
   \begin{figure*}
\centering 
  \includegraphics[width=18.cm]{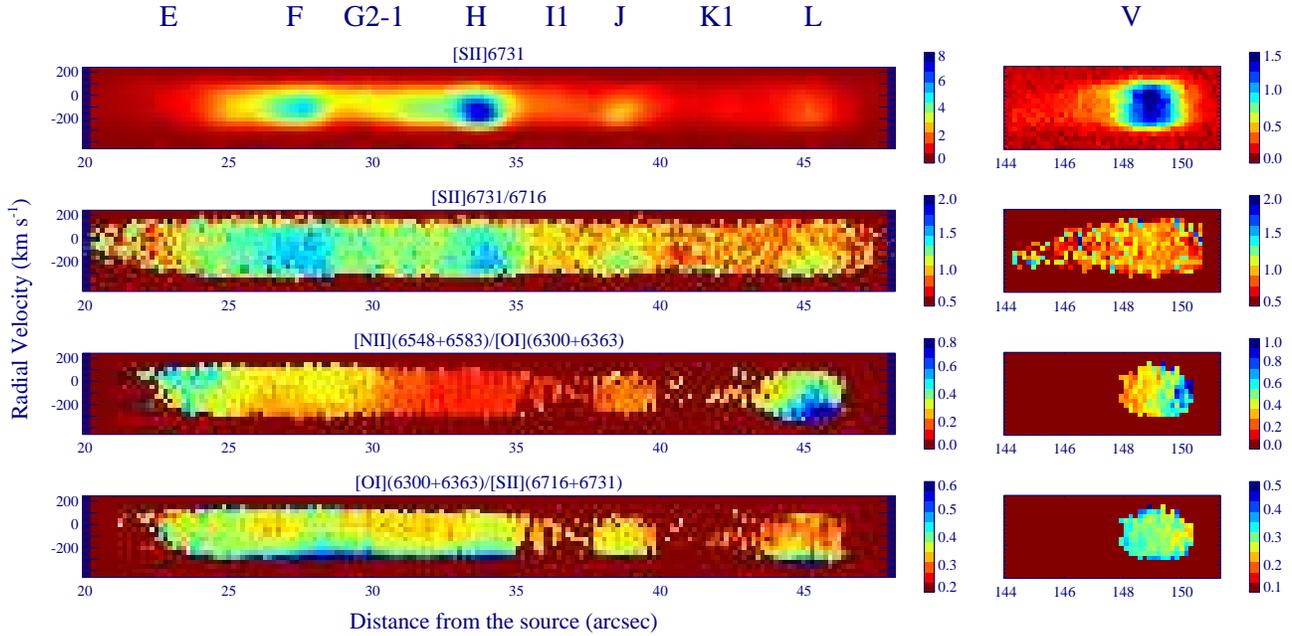}
   \caption{Position-Velocity diagrams of the forbidden line ratios used in 
the diagnostic technique for the HH 111 jet. 
{\em From top to bottom panel:} 
the \sii\lam6731 line intensity;
the \sii\lam6731/6716 ratio increases with electron density, \en;
the \nii/\oi\, ratio mainly depends on the ionisation fraction, \xe, and
increases for increasing \xe;
the \oi/\sii\, ratio increases for increasing temperature, \te.}\label{hh111_ratios}
    \end{figure*}
   \begin{figure}
\centering 
  \includegraphics[width=8.cm]{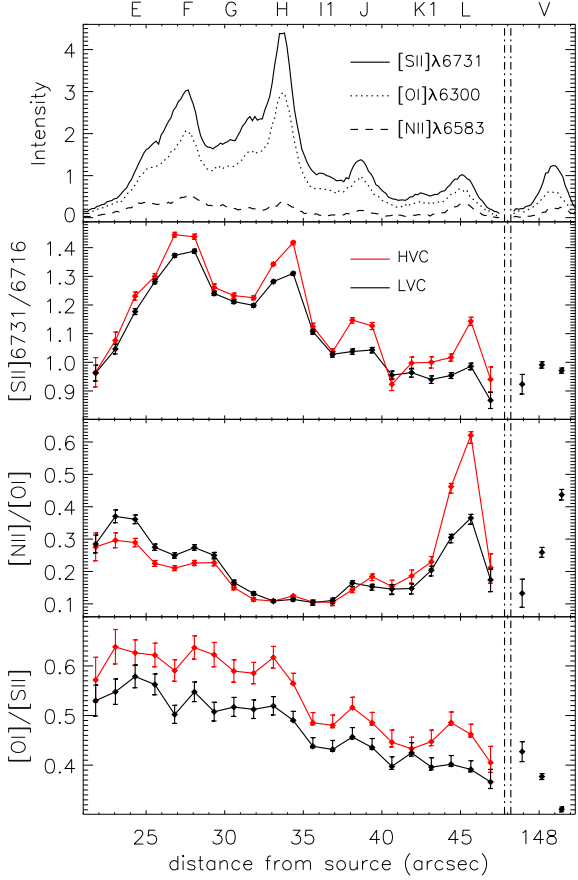}
   \caption{Variation of the forbidden line ratios used in 
the diagnostic technique for the HH 111 jet as a 
function of distance from the source and in two velocity intervals:
the High Velocity Interval (HVI, red points/line, $v$$<$-100 \kms) and 
the Low Velocity Interval (LVI, black points/line, $v$$>$-100 \kms). 
In contrast, in the terminal bow (knot V) the ratios has been computed
by integrating the fluxes over the line spectral profile thus there is no 
information on the velocity intervals.
{\em From top to bottom panel:} 
intensity profiles of the optical lines, \sii\lam6731/6716 ratios,
\nii/\oi\, ratios, and \oi/\sii\, ratios.}\label{hh111_ratios_der}
    \end{figure}
   \begin{figure}
     \centering
     \includegraphics[width=8.cm]{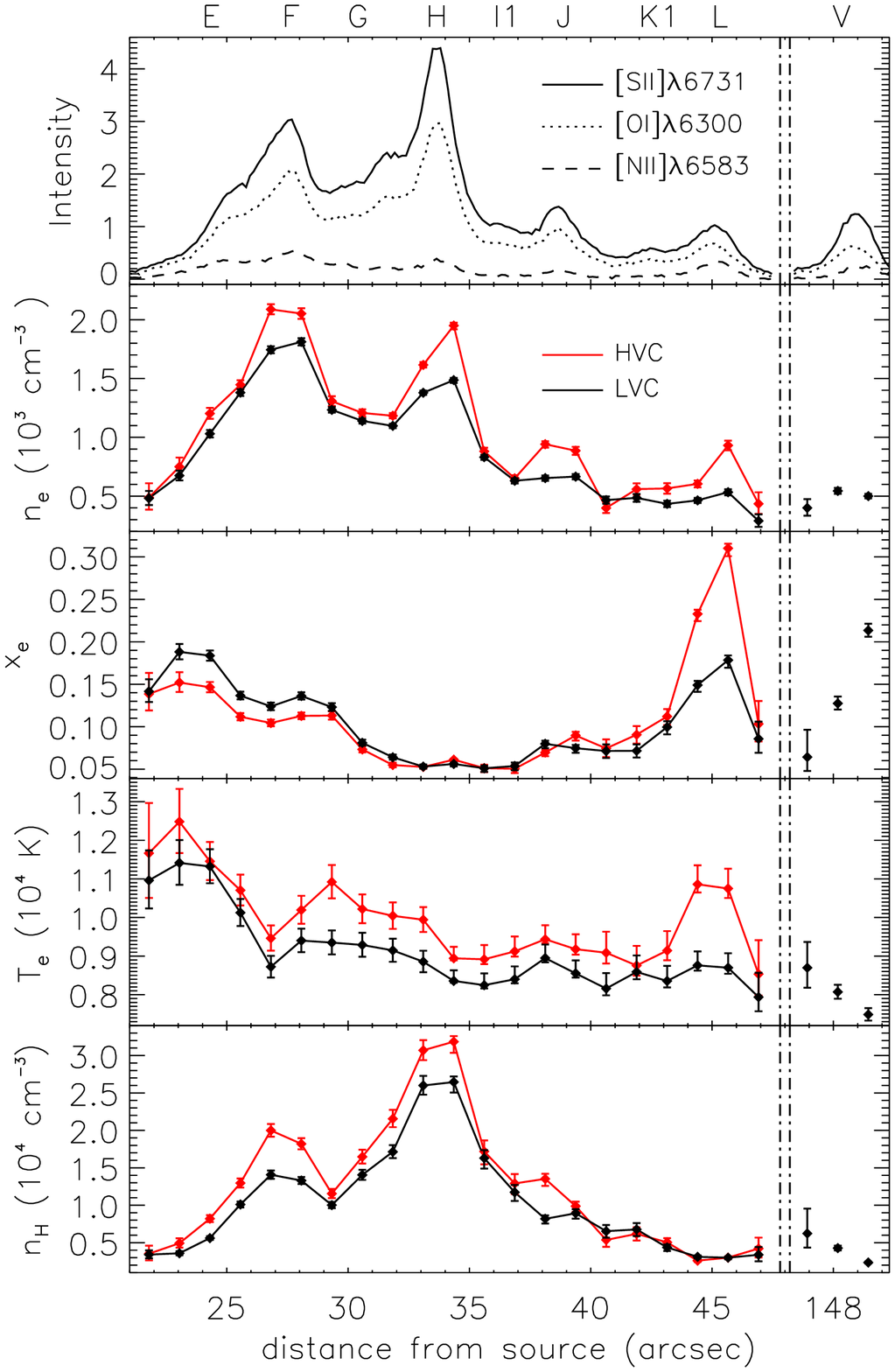}
   \caption{Variation of the physical parameters for the HH 111 jet as a 
function of distance from the source and in two velocity intervals:
the High Velocity Interval (HVI, red points/line, $v$$<$-100 \kms) and 
the Low Velocity Interval (LVI, black points/line, $v$$>$-100 \kms). 
In contrast, in the terminal bow (knot V) the parameters are derived 
integrating the fluxes over the line spectral profile thus there is no 
information on the velocity intervals.
{\em From top to bottom panel:} 
intensity profiles of the optical lines, the electron density, \en, in units
of 10$^3$ \cmc, the ionisation fraction, \xe, the temperature, \te, in units
of 10$^4$ K, and the total density, \nh, in units of 10$^4$ \cmc.
}\label{hh111_phys}
    \end{figure}


HH 111 is a well-known parsec-scale jet powered by the young star 
IRAS 05491+0247 located in the L1617 cloud in Orion (D=460 pc).
The blueshifted lobe is detected in the optical only from a distance of 
$\sim$10$''$ from the source. The first part of the beam is almost 
completely obscured by the molecular envelope, up to
$\sim$20$''$, where it emerges from the cloud showing
a chain of bright, equally spaced and well-collimated
knots up to a distance of $\sim$48$''$ (knots E-L, 
following the nomenclature from \citealt{reipurth97}).
These are followed by a series of more spaced and fainter knots up to 80$''$ 
(M, N, O, P), and then the jet terminates with a strong bow-shock located at 
$\sim$150$''$ (knot V).
The physical properties of HH 111 have been investigated by several authors
using both optical \citep[e.g., ][]{morse93,hartigan94} and near-infrared 
\citep[e.g., ][]{nisini02} lines.
\citet{reipurth97} and \citet{davis01}, instead, 
focused on the gas kinematics, 
through analysing \sii\, and \h\, line profiles, and
highlighted the presence of two velocity components peaking at 
$\sim$-75/-85 and $\sim$-15/-30 \kms.
In a previous work \citep{podio06} a detailed analysis of the HH 111 physical 
structure was performed using several spectral tracers, which showed
showing the density and temperature stratification in each spatially 
unresolved cooling region.
Nevertheless, none of the previous studies investigated the 
variation of the physical parameters as a function of the gas velocity, i.e.
in different velocity intervals.

As we explain in Sect.~\ref{sect:obs}  
the data of HH 111 show partially velocity resolved line profiles.
This is confirmed by the position-velocity (PV) diagrams 
of the computed line ratios in Fig.~\ref{hh111_ratios}, which clearly
show different value of the ratios in the two velocity intervals
in the jet beam (knots from E to L).
A comparison of the values of the ratios integrated over the two considered
velocity intervals (LVI and HVI) is also shown in 
Fig.~\ref{hh111_ratios_der}.
Since these ratios depend on the gas physical conditions,
they give a qualitative idea of the variations 
of electron density, temperature, and ionisation fraction along the jet 
and with the gas velocity.
The PV of the ratio between the sulphur doublet lines (\sii\lam6731/6716),
for example, shows that n$_{e}$ peaks at the positions of the brightest 
knots and at high velocities.   
The \nii/\oi\, and \oi/\sii\, ratios, on the other hand, 
indicate a peak of the excitation conditions (x$_{e}$, T$_{e}$)
in the high velocity interval of knot L.
Interestingly, HST high-angular resolution images of the jet 
clearly show that HH 111 is a chain of shock working surfaces
where knot L has the clearest ``bow'' morphology \citep{reipurth97} and 
is the knot with the largest \sii\, and \h\, line profiles 
\citep{reipurth97,davis01} and the largest shock velocity \citep{hartigan01}
in the jet beam.
Our PV diagrams confirm this picture highlighting 
the shock velocity structure.


By integrating the line fluxes over the seeing FWHM and over the LVI and HVI 
and then by applying the BE technique we obtain 
the variations of the physical parameters both along the jet
and in the two velocity intervals for knots E-L (see Fig.~\ref{hh111_phys}).
The derived parameters are in agreement with previous estimates 
\citep{podio06}:
n$_{e}$ varies between 10$^2$ and $\sim$2.2~10$^3$~cm$^{-3}$, x$_{e}$ goes 
from a few percent to $\sim$0.3, T$_{e}$$\sim$0.7-1.3~10$^4$~K, 
and n$_{H}$ ranges between $\sim$2~10$^3$ and $\sim$3~10$^4$~cm$^{-3}$. 
With respect to previous analyses, however, 
the increased spatial sampling (more than twice 
with respect to previous studies) and the 
velocity information highlight several shocks features, 
such as n$_{e}$ maxima 
in the HVI of the brightest knots, higher values of T$_{e}$ in the HVI
along the full jet length, and, finally, 
a strong and steep increase of x$_{e}$ and T$_{e}$
in the HVI of knot L.
Note that very weak emission in the \oiii\lam\lam4959,5007 lines was
detected by \citet{morse93} in this knot. The presence of oxygen in 
\oiii\, form is not accounted for in the adopted diagnostic technique.
Nevertheless, since the emission in these lines is very weak, we
tentatively applied the diagnostic to knot L and 
the obtained high values of \xe\, and \te\, are
in agreement with the detection of such high-excitation lines.  
These results are an important observational validation of the shock 
velocity structure predicted by models \citep{hartigan87}.

We do not apply the diagnostic to knots M, N, O, and P because of the
very weak emission in the \nii\, lines.
Nevertheless, the electron density can be inferred by integrating the \sii\, 
emission over the knot spatial profile to increase the S/N, 
which is low at the position of these faint knots.
We obtain \en\, values of about 150 \cmc, 200 \cmc, 250 \cmc, and 150 \cmc\, 
in the HVI of knot M, N, O, and P and values of $\sim$100 \cmc\, lower in the
LVI for the same knots.  
The average values are in agreement with previous results from \citet{podio06}. 

Previous works did not detect high excitation lines at the position 
of the terminal bow shock (knot V) \citep{morse93b}, 
thus we apply our diagnostic to this knot.
Even if two velocity component has been detected in the \sii\, lines by
\citet{reipurth97} the PVs in Fig.~\ref{hh111_ratios} 
do not show different values of the ratios in the two velocity intervals.
This can be due to the low spectral resolution combined with the much
lower intensity of knot V with respect to the knots in the jet beam (knot E-L). 
Thus we integrate the line fluxes over their
spectral profiles obtaining the trend shown in Fig.~\ref{hh111_phys}.
The electron and total density are of $\sim$400-550 \cmc\, and 
$\sim$2-5 10$^3$ \cmc, respectively.
The ionisation fraction and the temperature show an unexpected 
anti-correlation and varies between $\sim$0.23 and $\sim$0.07, 
and $\sim$0.75 10$^4$ K and $\sim$0.85 10$^4$ K moving from the shock front 
toward the source.
Note that despite the high shock velocity of knot V \citep{hartigan01} we did 
not infer high values of \xe\, and \te\, as for knot L.

\subsection{The Sz68 jet}
\label{sect:sz68}
   \begin{figure}
     \centering
     \includegraphics[width=8.cm]{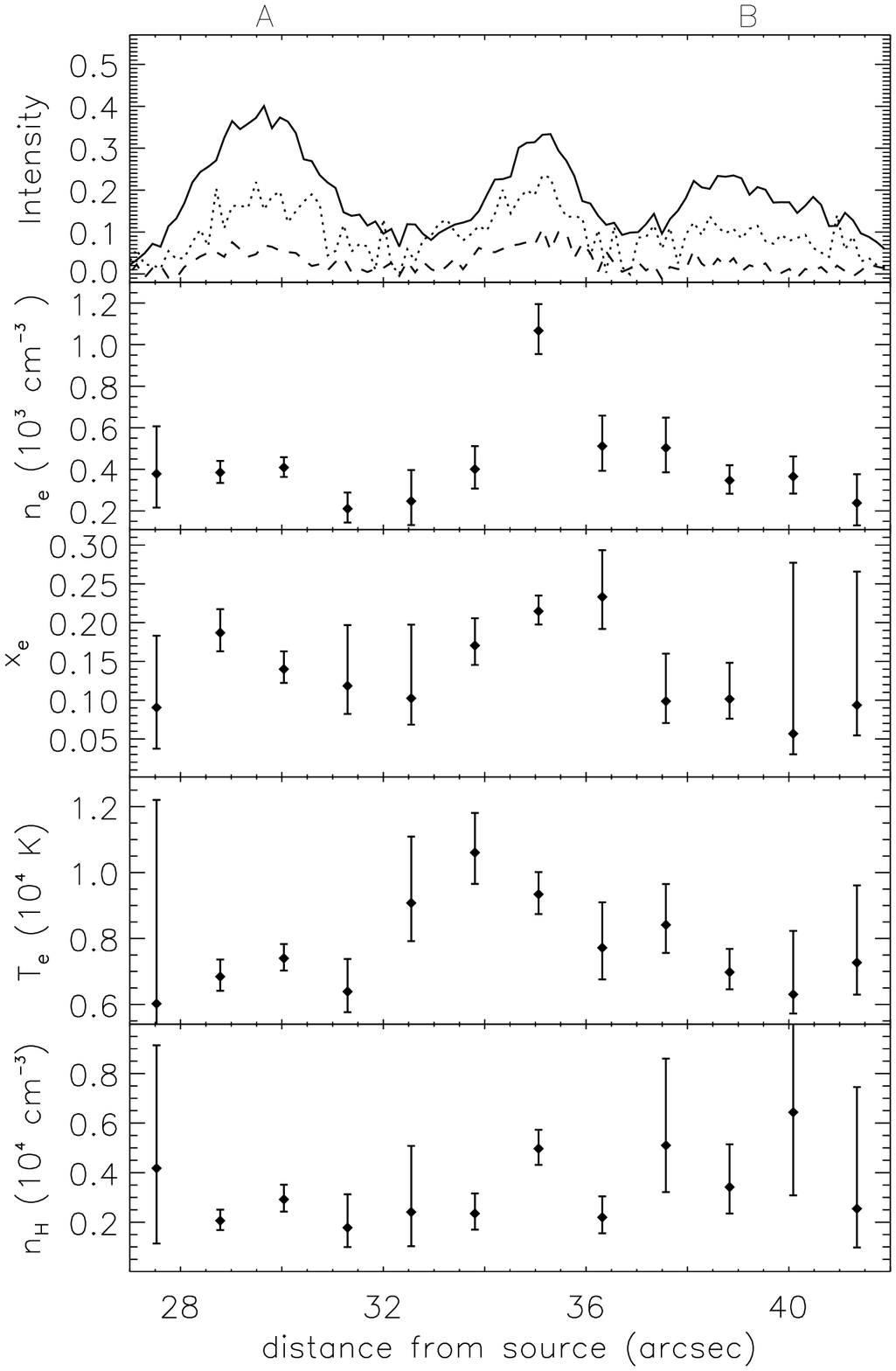}
   \caption{Variation of the physical parameters for the Sz68 jet as a 
function of the distance from the source.
{\em From top to bottom panel:} 
intensity profiles of the optical lines, the electron density, \en, in units
of 10$^3$ \cmc, the ionisation fraction, \xe, the temperature, \te, in units
of 10$^4$ K, and the total density, \nh, in units of 10$^4$ \cmc.}\label{sz68_phys}
    \end{figure}


The Sz68 jet has never been analysed through spectral diagnostic techniques.
This jet is located in the B228 molecular cloud between the two 
emission-line stars Sz68 and Sz69.
The collimated chain of \sii\, shock-excited blueshifted 
emission extending  up to 34$''$ 
from Sz68 at a position angle of 135\degr\, was detected by \citet{heyer89}.
They report three knots located at 21\arcsec, 28\arcsec, 
and 34\arcsec\, from the source, the brightest ones being the first knot, A, 
and the last one, B.
Comparing the position of the knots in the spectra of \citet{heyer89} with
the positions in our EFOSC2 data (see upper panel of Fig.~\ref{sz68_phys} for 
the knots spatial profile), we derive an estimate of the knot proper motions.
This indicates a tangential velocity of $\sim$600 \kms\, for knot A, $\sim$460 \kms\,
for the intermediate knot, and $\sim$330 \kms\, for knot B,
which combined with the radial velocity 
corrected for the source heliocentric velocity 
($\sim$-55 \kms\, for knot A and $\sim$-29 \kms\, for knot B, 
\citealt{heyer89}) 
gives an upper limit for the inclination angle out of the plane of the sky 
of $\sim$5\degr.

We applied the BE technique to the line fluxes measured along the three knots
by integrating all over the line spectral profiles.
Since this object is very faint, the error-bars are much larger than for the 
other jets.  
The inferred electron density is quite low varying between 200~\cmc\, and 
1.1~10$^3$~\cmc\,. These values are slightly lower with respect to those 
derived by \citet{heyer89}.
The ionisation fraction is 0.05-0.25 and the temperature ranges between 
0.6-1.1~10$^4$ K, with a maximum at the position of the intermediate knot,
where there is a peak of \en, as well.
Finally the hydrogen total density is 2-6~10$^3$ \cmc. 

The determination of the jet physical conditions, as well as 
the jet tangential velocity
and inclination angle allows us to estimate the mass loss rate as
$\dot{M}_{\rm jet}$ = $\mu$\,$m_{H}$\,$n_{H}$\,$\pi$\,$r_{J}$$^2$\,$v_J$,
where $\mu$=1.41 is the mean atomic weight, $m_{H}$ the proton mass,  $n_{H}$
the hydrogen density and $r_J$ and $v_J$, respectively, the jet radius and 
velocity.
By assuming a typical jet radius of $\sim$100-300 AU we obtain a mass loss rate
between a few 10$^{-8}$ and a few 10$^{-7}$\msolyr.
These values are of the same order of magnitude of \mjet\, found for typical
HH jets \citep{hartigan94,bacciotti99,podio06}.

\section{Dust Reprocessing in the Jets}
\label{sect:dust}

Refractory species, such as Ca, Fe, Ni, Cr, Ti, Mg, Si
are often depleted in the interstellar 
medium because their atoms are locked onto dust grains \citep{savage96}. 
The calcium gas-phase abundance, for example, 
is estimated to be $\sim$2 10$^{-8}$ in Orion \citep{baldwin91}, 
i.e. two orders of magnitude lower 
than its Solar abundance ($\sim$2.04 10$^{-6}$, \citealt{asplund05}),
and up to 4 order of magnitude lower than Solar in $\zeta$ Oph 
\citep{savage96}. 
On the other hand, the passage of shocks can partially or completely
destroy the dust releasing the refractory atoms into 
the gas cloud.

Theoretical models of dust reprocessing in shocks, such as those driven by 
supernova explosions or those occurring in stellar jets, estimated the dust 
destruction rate by considering the different processes at work, i.e. thermal 
and inertial sputtering, photoevaporation and shattering. 
They showed that the relative importance of these processes, and thus the 
efficiency of the shock in reprocessing the dust, strongly depend on a 
number of parameters such as the shock velocity, the gas pre-shock density, 
the intensity of the magnetic field, the size and 
structure of the dust grains 
\citep[e.g., ][]{jones94,jones00,draine03,guillet09}.

From an observational point of view an estimate of the depletion of refractory 
species with respect to their Solar abundance can be used to gauge the amount 
of dust grains in stellar jet beams and thus to evaluate the efficiency of 
these mild shocks in destroying the dust. 
This in turn can help constraining the location and size of the region
around the star from where the jet originates. 

In Sect.~\ref{sect:method} we explain in detail the method used
to estimate the abundance of calcium in the gas-phase, while
in  Sect.~\ref{sect:dust_results} we present the results obtained and
in Sect.~\ref{sect:dust_discussion} we
compare them with the predictions of theoretical models.

\subsection{Observational estimates of calcium gas-phase abundance in stellar 
jets: the method}
\label{sect:method}

To estimate the depletion of a given refractory specie, 
in our case calcium,
with respect to its Solar abundance, we compare ``expected'' and 
observed ratios between refractory and non-refractory species
known to be present in Solar abundance in the ISM (e.g. \citealt{savage96}).

The ``expected'' ratios can be computed through two different methods:
\begin{itemize}

\item[(i)] by using shock models which predict the gas physical conditions and 
thus the emission in different lines behind the shock front. This method was
used by \citet{mouri00} and, in part, by  \citet{nisini02} to estimate
Fe gas-phase abundance in HH jets. They found that Fe is strongly depleted
in these objects, up to 80\% with respect to its Solar abundance;\\

\item[(ii)] by using the gas physical conditions inferred through spectral 
diagnostic tools. 
In this case the estimate does not depend upon assumptions of the
shock characteristics (velocity, pre-shock density, magnetic field intensity)
but depends on the accuracy of the determinations of the gas parameters.
\citet{beck-winchatz94,beck-winchatz96} used this method 
to estimate Fe and Ni gas-phase abundance in HH jets. 
They found that Fe is not, 
or only partially, depleted in HH jets, while Ni is surprisingly over-abundant.
The method was then used by \citet{nisini02,nisini05} and \citet{podio06} to 
estimate the gas-phase abundance of both Fe and Ca obtaining
a strong depletion (up to 90\%) of these species which, however, may vary 
along the jet and decreases down to 0\% in some knots.
\end{itemize}

The main assumption of method (ii) is that the considered 
lines from refractory and non-refractory species are emitted in the same
region behind the shock front, in such a way that 
the filling factor and the gas physical conditions are the same.
This has been verified by \citet{nisini05} (see their Fig.~3 and corresponding
discussion).

Following \citet{nisini05} and \citet{podio06}, here we use the 
\caii\lam7291/\sii\lam6731 
ratio to estimate the Ca gas-phase abundance in our sample of HH jets.
These lines have similar excitation temperatures ($\sim$2\,10$^4$ K for 
\caii\lam7291 and $\sim$2.2\,10$^4$ K for \sii\lam6731) and, despite the 
critical density of the calcium line being larger than that of the
sulphur line (n$_{cr}$(\sii)$\sim$2.5\,10$^4$ \cmc, 
n$_{cr}$(\caii)$\sim$5\,10$^7$ \cmc), at the typical low density of HH 
jets the emission of the two lines peaks at the same distance from the shock 
front. 
In the same region the \sii, \oi, and \nii\, lines which are used
to derive the gas physical conditions along the jet are also excited.
Thus the estimated parameters can be used consistently 
to model the theoretical \caii/\sii\, ratio.

Following \citet{hartigan04} we assume that the \caii\, lines are 
collisionally excited and compute the level populations. 
As in previous works \citep{nisini05,podio06} we assume that there is no 
calcium in the
form of Ca$^0$, because its ionization potential is very low, $\sim$6.1\,eV.
On the other hand, the ionization potential of \ioncaii\, is also quite low,
$\sim$11.9\,eV, i.e. lower than that for hydrogen ($\sim$13.6\,eV).
In \citet{nisini05} and \citet{podio06} 
we computed the \caii/\sii\, ratios by assuming
that all calcium is in the form of Ca$^{+}$.
This is a reasonable approximation in the case of poorly ionized objects
such as HH 1 and HH 111 (\xe$\le$0.3).
However, the presence of Ca in the form of Ca$^{++}$ cannot be neglected when
analysing higher excitation jets such as HH 83 and HH 24.
The coefficients for collisional ionization and radiative and dielectronic
recombination of Ca$^{+}$ and H$^{0}$ are very similar for temperatures 
between 8\,10$^3$ K and 3\,10$^4$ K 
(at T$\sim$2 10$^4$ K, the recombination rates for Ca$^{+}$ and H$^{0}$ are:
$\alpha$(Ca$^+$)$\sim$3.9 10$^{-13}$ cm$^3$\,s$^{-1}$, 
$\alpha$(H$^0$)$\sim$2.5 10$^{-13}$ cm$^3$\,s$^{-1}$; 
while the coefficients for collisional ionization are:
C(Ca$^+$)$\sim$1.2 10$^{-11}$ cm$^3$\,s$^{-1}$, 
C(H$^0$)$\sim$1.6 10$^{-11}$ cm$^3$\,s$^{-1}$). 
Therefore, considered that at these temperatures all calcium is in the
form of Ca$^{+}$ and Ca$^{++}$, in conditions of ionization equilibrium the 
ionization fraction of Ca$^{+}$ (defined as x(Ca$^{+}$)=Ca$^{++}$/Ca) and 
H$^{0}$ are almost the same: x(Ca$^{+}$)$\sim$\xe.
In stellar jets, however, the ionization fraction of hydrogen is never at 
equilibrium with the local temperature because,
as explained in \citet{bacciotti99}, the gas is moving along the jet and
the recombination timescale at the observed electron densities is of 
$\sim$10$^3$ yr, i.e. of the same order of the jet dynamical time.
Because of the similarity of the recombination and collisional ionization
coefficients, however, Ca$^{+}$ and H$^{0}$ atoms can be thought to be 
undergoing the same processes when moving along the jet.
Thus we assume that, at every position along the jet, the  Ca$^{+}$ ionization
fraction is equal to the hydrogen ionization fraction, \xe.
Taking Ca and S gas-phase abundances equal to Solar ones
\citep{asplund05}, we can finally compute the \caii/\sii\, theoretical ratios.

Note that,
in contrast with previous works \citep{nisini05,podio06}, here we use the
\caii\lam7291/\sii\lam6731 ratio, instead of the 
\caii\lam(7291+7324)/\sii\lam(6716+6731), 
since the \caii\lam7324 line may be blended with 
the \oii\lam\lam7318.6,7319.4 and the \oii\lam\lam7329.9,7330.7 lines.
As explained in \citet{hartigan04} the calcium line ratio 
\caii\lam7291/\lam7324 is expected to be $\sim$1.5 both in the case of
collisional excitation and for fluorescent pumping from the
ground state.
Thus by measuring the ratio between the calcium lines we can check for the
presence, if any, of \oii\, emission blended with the \caii\lam7324 line.
We find that this is not the case for the HH 1 and the HH 34 jets 
\citep{nisini05,podio06}, but may be the case in the HH 111, HH 83 and HH 24 
jets which show higher excitation conditions in some of the knots 
(see Figs.~\ref{hh111_phys}, \ref{hh83_phys}, and \ref{hh24_phys}).
These jets, in fact, show a wider \caii\lam7324 line in some knots 
indicating blending with the \oii\, lines.
Moreover, in our EFOSC2 spectra, the \caii\lam7324 line is at the border of the
detector and hence the line flux can be affected by distortion and/or higher 
S/N.
Thus we decided to use the \caii\lam7291/\sii\lam6731 for all the jets in our 
sample.

For the HH 111 and the HH 1 jet we corrected the observed \caii/\sii\,
ratios for reddening following the procedure explained in 
Sect.~\ref{sect:diag}.
For the HH 83 and the HH 24 jet we have no estimates of the visual extinction, 
thus we could not correct the Ca$^{+}$ and S$^{+}$  line fluxes for 
reddening and the observed ratios (computed assuming  A$_{\rm V}$=0)
are actually upper limits to the real values 
(the latter decrease for increasing A$_{\rm V}$). 
This means that the calcium depletion inferred from  
Figs.~\ref{hh83_depl} and \ref{hh24_depl} is a lower limit. 

The errors affecting the observed ratios are due to the signal-to-noise
over the measured line fluxes.
For HH 1 and HH 111 we consider also the errors due to the uncertainty in 
estimating A$_V$.

The errors on the theoretical ratios are due to the uncertainty of the estimated
physical parameters and are computed by varying \en, \xe, and \te\, between
their minimum and maximum values.
The influence of these errors is represented in the form of a green stripe in 
Figs.~\ref{hh111_depl}, \ref{hh1_depl}, \ref{hh83_depl}, and \ref{hh24_depl}.
Since, as explained in Sect.~\ref{sect:diag}, the choice of the abundance set
may affect the absolute values of \xe\, and \te, we also computed the 
variations of \caii/\sii\, when adopting the alternative abundance set 
estimated by \citet{esteban04}. 
The difference between the \caii/\sii\, ratios obtained with the two sets
is illustrated by the width of the beige stripes in         
Figs.~\ref{hh111_depl}, \ref{hh1_depl}, \ref{hh83_depl}, and \ref{hh24_depl}.

\subsection{Results: Ca depletion along HH jets.}
\label{sect:dust_results}

   \begin{figure}
     \centering
     \includegraphics[width=8.cm]{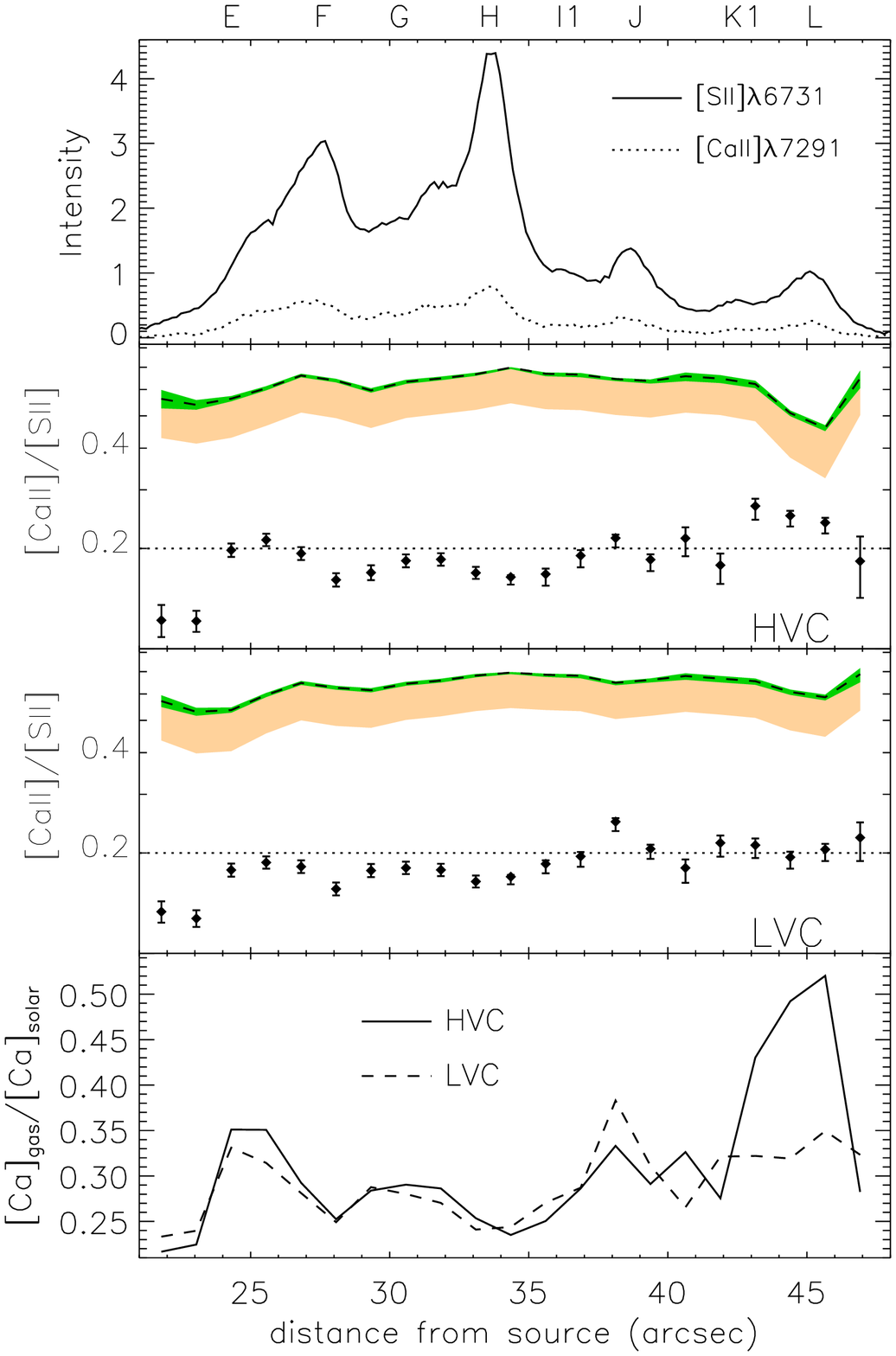}
   \caption{Comparison between observed (diamonds) and predicted (dashed lines)
\caii\lam7291/\sii\lam6731 along the HH 111 jet and in two 
velocity intervals: the HVI ($v$$<$-100 \kms, top panel) and 
the LVI ($v$$>$-100 \kms, bottom panel).
The dotted line refers to the predicted ratios computed assuming Solar
abundances from \citet{asplund05} and the green stripe shows the variation of 
the ratios due to the errors on the inferred physical parameters. 
The beige stripe, on the contrary, shows the variations of the predicted 
ratios when assuming non-solar abundances from \citet{esteban04}. 
In any case, 
the difference between observed and predicted ratios indicates a depletion 
of the Ca gas-phase abundance, which is quantified in the bottom panel
using the predicted ratios computed assuming Solar
abundances (dotted lines).}
\label{hh111_depl}
    \end{figure}
   \begin{figure}
     \centering
     \includegraphics[width=8.cm]{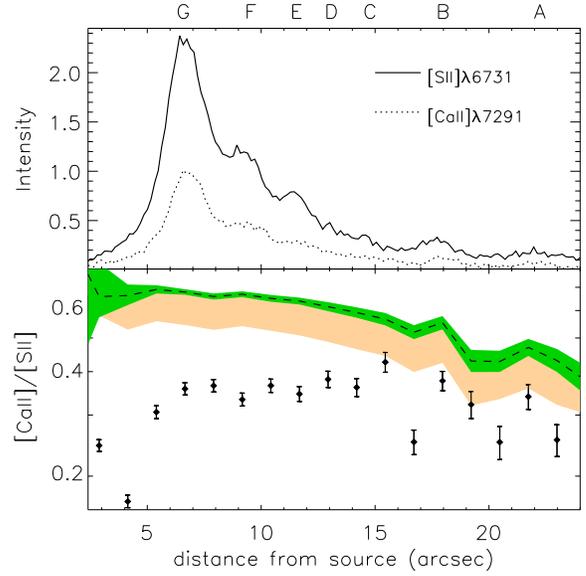}
   \caption{Comparison between observed (diamonds) and predicted (dashed line)
\caii\lam7291/\sii\lam6731 along the HH 1 jet.
The green stripe and the beige stripe show the variation of the predicted 
ratios, as explained in Fig.~\ref{hh111_depl}. 
The difference between observed and predicted ratios indicates a depletion of 
the Ca gas-phase abundance.}\label{hh1_depl}
    \end{figure}
   \begin{figure}
     \centering
     \includegraphics[width=8.cm]{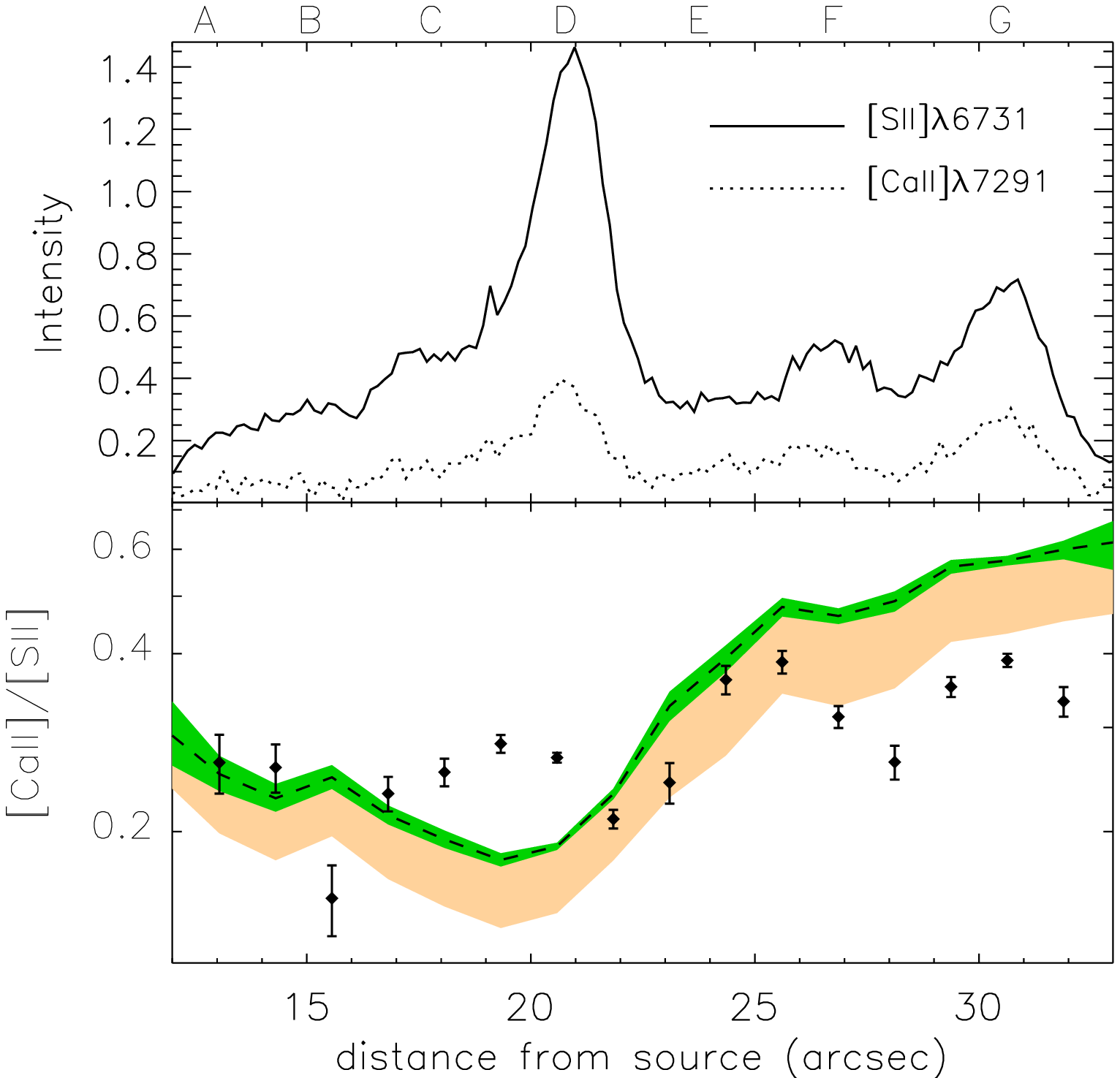}
   \caption{Comparison between observed (diamonds) and predicted (dashed line)
\caii\lam7291/\sii\lam6731 along the HH 83 jet.
The observed ratios are not corrected for reddening (A$_{\rm V}$=0), and
are thus upper limits.
The green stripe and the beige stripe show the variation of the predicted 
ratios, as explained in Fig.~\ref{hh111_depl}. 
}\label{hh83_depl}
    \end{figure}
   \begin{figure}
     \centering
     \includegraphics[width=8.cm]{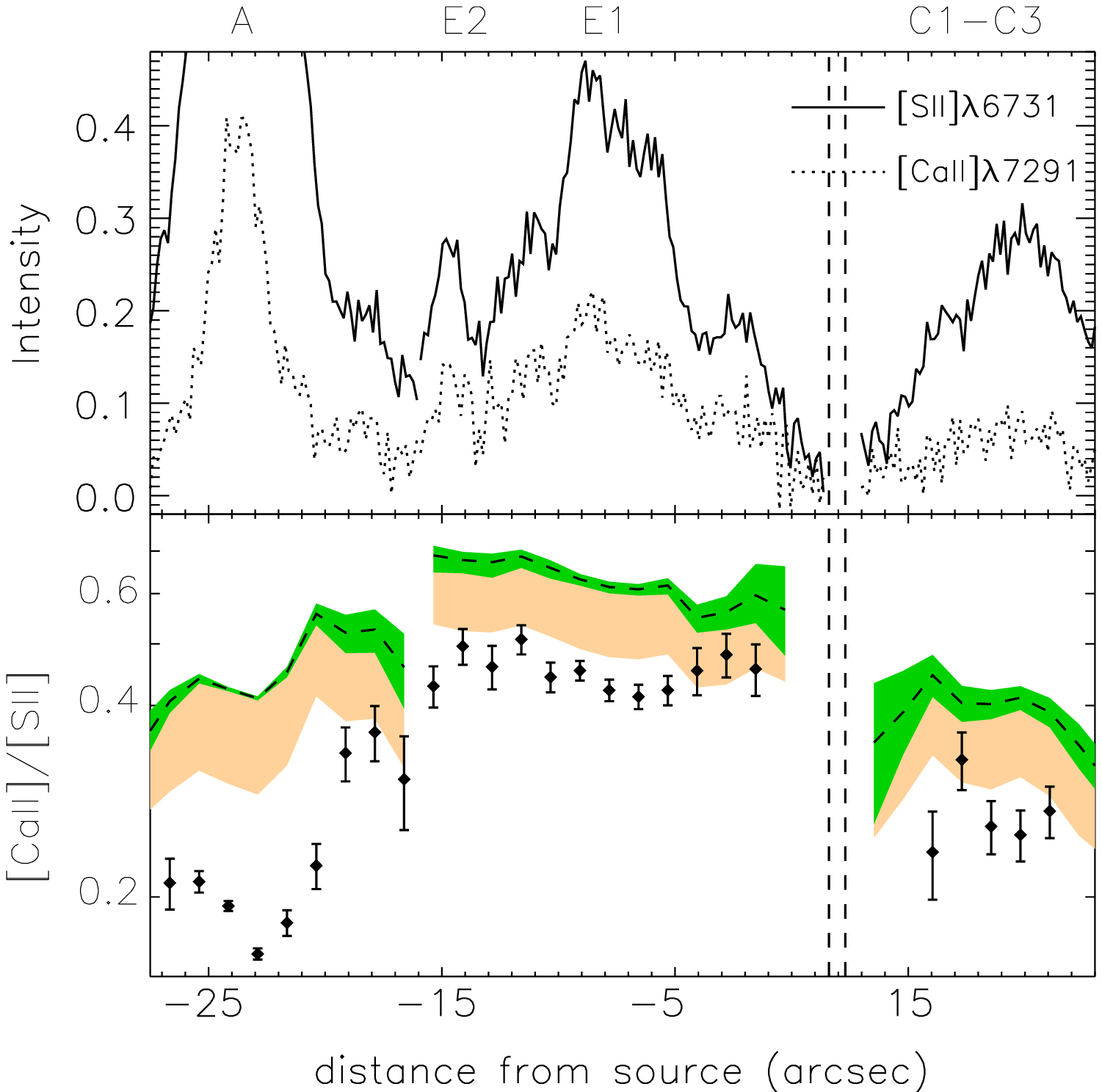}
   \caption{Comparison between observed (diamonds) and predicted (dashed line)
\caii\lam7291/\sii\lam6731 along the HH 24 jet.
The observed ratios are not corrected for reddening (A$_{\rm V}$=0) and
thus they are an upper limits.
The green stripe and the beige stripe show the variation of the predicted 
ratios, as explained in Fig.~\ref{hh111_depl}. 
}\label{hh24_depl}
    \end{figure}
\begin{table}
\caption[]{\label{tab:depletion} Gas-phase abundance of calcium
with respect to Solar Abundances.}

\vspace{0.5cm}
    \begin{tabular}[h]{cc}
      \hline 
Jet          & [Ca]$_{gas}$/[Ca]$_{Solar}$$^{a}$  \\
\hline   
HH 111       & 0.21-0.52   \\
HH 1         & 0.25-0.79 \\
HH 83        & 0.53-1.57$^{*}$ \\
HH 24 A      & 0.40-0.69$^{*}$ \\
HH 24 E      & 0.62-0.85$^{*}$ \\
HH 24 C      & 0.52-0.82$^{*}$ \\
HH 34$^{b}$  & 0.3-1 \\
\hline 
Orion Cloud$^{c}$  & 0.01 \\
\hline 
      \end{tabular}

~$^{a}$(Ca/H) Solar abundance from \citet{asplund05}: 2.04 10$^{-6}$ \\
~$^{b}$from \citet{podio06} \\
~$^{c}$(Ca/H) abundance estimated in the interstellar medium in Orion by 
\citet{baldwin91}: 2 10$^{-8}$ \\
~$^{*}$the values of [Ca]$_{gas}$/[Ca]$_{Solar}$ for HH 83 and HH 24 have
been estimated assuming A$_{\rm V}$=0, thus they are upper limits to the real 
values.   

\end{table}

The comparison between the observed and the predicted \caii/\sii\, ratios
is shown in Figs.~\ref{hh111_depl}, \ref{hh1_depl}, \ref{hh83_depl}, 
and \ref{hh24_depl} and the inferred values of the calcium gas-phase abundance  
are summarized in Table \ref{tab:depletion}.
Note that [Ca]$_{gas}$/[Ca]$_{Solar}$ values in Table \ref{tab:depletion}
and in the bottom panel of Fig.~\ref{hh111_depl} are derived by using the 
\caii/\sii\, ratios predicted assuming Solar elemental abundances from 
\citet{asplund05} (illustrated by a dotted line in the figures).

Along the HH 111 and HH 1 jets the observed \caii\lam7291/\sii\lam6731 ratios 
are lower than the predicted ones, i.e. calcium is depleted 
with respect to its Solar abundance, the depletion being between
20\% and 80\%. 
For these jets the results agree with what has been found in \citet{nisini05} 
and \citet{podio06}, the
slight difference being explained by the fact that in those works
the presence of calcium in the form of Ca$^{++}$ was neglected and a
different set of abundances was assumed.

On the contrary, along HH 83 the dust grains appear to have been destroyed
substantially, as the derived Ca abundance is equal or even larger than Solar 
in the first knots and is depleted up to 47\% at distances larger than 
25\arcsec\, from the source. 
In the case of HH 83, however, we did not correct the fluxes for
extinction, thus the obtained values of Ca gas-phase abundance are upper 
limits. 
It can be shown that with an A$_{\rm V}$=5 the Ca gas-phase abundance is between
0.36 and 1.
In any case the low depletion in this jet appear to be consistent with the
jet high excitation. 

Finally, in HH 24 we derive the \caii\lam7291/\sii\lam6731 ratio only in 
a few of the observed knots because for the fainter ones the Ca$^{+}$ emission 
is below our assumed S/N threshold (equal to three times the background noise).
Fig.~\ref{hh24_depl} shows that the depletion 
along the jet, show no evident trend: calcium is depleted by 48\%-18\%
in HH 24 C, by $\sim$38\%-15\% in HH 24 E and by $\sim$60\%-31\% in HH 24 A.
The lack of a trend along the jet reflects the situation found for the
physical parameters (see Fig.~\ref{hh24_phys}) and supports the idea that
HH 24 C, E, and A may not belong to a single HH jet \citep{eisloffel97}. 
Note that also in the case of HH 24 
we did not correct the observed \caii/\sii\,
ratios for extinction thus the inferred Ca depletion is a lower limit.

The width of the beige stripes in 
Figs.~\ref{hh111_depl}, \ref{hh1_depl}, \ref{hh83_depl}, 
and \ref{hh24_depl} show the variation of the theoretical \caii/\sii\, 
ratios when adopting abundances from \citet{esteban04}.
Note that also in this case
calcium turns out to be largely depleted in most of the knots, 
although the inferred Ca depletion is lower.

\subsection{Discussion.}
\label{sect:dust_discussion}

In general, we find an amount of Ca depletion  
comparable to the Fe depletion estimated in other works \citep{beck-winchatz96,bohm01,mouri00,nisini02,nisini05,podio06,garcialopez09}. 
Together the results obtained for the two different refractory species 
indicate that, despite the presence of shocks, 
there are still dust grains in the jets. 
However, partial or total destruction of dust grains
can be inferred by comparing the level of calcium depletion in the jets with 
that of the diffuse interstellar medium in Orion 
(see Tab.~\ref{tab:depletion}).


According to the most recent models of dust reprocessing from 
\citet{guillet09} the dust destruction rate in J-shocks with velocities lower 
than 50 km s$^{-1}$ and pre-shock densities from 10$^4$ to 10$^6$ \cmc\,
is around a few percent.
Previous models by \citet{jones94} predict a destruction rate for silicate
grains as high as 50\% for shock velocities of $\sim$100 km s$^{-1}$ and low
values of the $b$ parameter ($b=B_{0} n_{0}^{-1/2}$, where $B_{0}$ is the 
intensity of the magnetic field and $n_{0}$ is the pre-shock density).
Finally, up to 90\% of Fe and 60\%-70\% of Mg and Si can be released in
gaseous form in C-shocks with velocities of only $\sim$45 km s$^{-1}$
\citep{may00}.  
It is important to stress that none of the above mentioned models has been 
tested with grains containing atoms of Ca and that
these predictions are highly dependent on dust size,
structure, and composition. 
Different results, for example, are retrieved when considering
silicate or graphite grains, and if the grains are solid or porous
\citep{jones94,guillet09}.

In any case, none of the present models predict a total dust grain 
destruction
in the mild shocks that occurr in the jet beam where the material interacts
with gas already put into motion by the passage of previous fronts 
($v_s$ $\le$100 km s$^{-1}$ in the jet beam, \citealt{hartigan01}).
This discrepancy between the predictions of shock models and the measured
Ca gas-phase abundance is easily solved if we assume that the dust has been
reprocessed in the disk or circumstellar region 
before being extracted and accelerated into jets.
According to recent theoretical models and observations, the accretion
disk in Herbig and Classical T Tauri Stars is populated by dust grains only 
beyond the so-called dust evaporation radius, R$_{evp}$ 
\citep[e.g., ][]{isella05}. 
One can thus infer that 
the material in the jet which is coming from a region smaller than 
R$_{evp}$ would not contain grains, as these have been 
destroyed by the stellar radiation.
The dust evaporation radius is located at $\sim$0.1-1.5 AU in Herbig 
stars \citep{eisner07} and at $\sim$0.05-0.3 AU in T Tauri stars
\citep{akeson05}, while there are no estimates in younger and more embedded 
sources such as the emitting sources of the HH jets in our sample. 

On the other hand, the depletion is observed at various levels in the beam,
and to explain the presence of dust grains
we have to assume that part of the material in the jet is coming from a region 
of the disk which extends beyond  R$_{evp}$.
Note that  we are not observing ambient dust since this has been swept-out or  
destroyed by the passage of the leading bow-shock, which has larger 
shock velocities ($v_s$ up to $\sim$150 \kms\, in the bow of HH 1, 
\citealt{bally02}) and the dust reformation time-scale is much larger 
than the dynamical timescale of the jet, which is of 10$^3$-10$^4$ yr. 
Moreover, the rise in gas-phase calcium with distance from the source
along the HH 111 and HH 1 jets argues 
against major entrainment of matter along the jet, as would be suggested by 
models of prompt and/or turbulent entrainment \citep{masson93,stahler94}.

Interestingly, in HH 111, for which we could perform the analysis in
the two velocity intervals, the Ca depletion is at a minimum in  
the HVI of the bow-shaped knot L 
(see HST images from \citealt{reipurth97}) which is also the one with the
higher shock velocity ($v_s$$\sim$60 \kms, \citealt{hartigan01}).
This result suggests that the efficiency of shocks in destroying dust grains
strongly depends on the shock velocity, in agreement with the 
predictions of theoretical models \citep{jones94,draine03,guillet09}.
Our estimate also indicates that this shock should be able to destroy at 
least 40\% of the dust grains present at this position.

The proposed proportionality of destruction efficiency and shock velocity
is also supported by our finding 
that calcium is less or not depleted along HH 83 and HH 24.
These jets, in fact, show higher excitation conditions in comparison to HH 1 
and HH 111 suggesting that their gas has undergone stronger internal shocks.



\section{Summary and Conclusions}
\label{sect:concl}

In this paper we analyse partially velocity resolved optical spectra of a
sample of HH jets located in the Orion and in the Lupus Molecular Cloud
(HH 111, HH 1/2, HH 83, HH 24 M/A/E/C, and Sz68), to investigate
primarily the content of Ca in gas-phase and, in turn, the dust reprocessing
in jets.
To this aim we first apply 
spectral diagnostics \citep{bacciotti99} to derive the physical
conditions of the gas in the jets from the line
ratios of the detected forbidden lines, i.e. the values of the
electron and total density, the ionisation fraction, and the temperature.
These parameters are then used to estimate the expected Ca abundance.

We find that in the studied jets the electron density ranges between 50 and 
4~10$^3$ \cmc. 
The ionisation fraction varies between a few percent in low excited knots
where the \nii\, emission is very faint and $\sim$0.4-0.7 in high
excitation knots with strong \nii\, emission.
The temperature follows the same trend as the ionisation fraction, being
$\sim$10$^4$ K in the weakly ionised knots and $\sim$1.5-3 10$^4$ K in the
highly ionised ones, in agreement with predictions from shock
models \citep[e.g., ][]{hartigan87}.
Finally, 
the total density derived by combining the retrieved values of \en\, and \xe\,
varies between $\sim$100 and 6 10$^4$ \cmc.

The most interesting results are obtained for the HH 111 jet for which we derive
the values of the physical parameters separately in two velocity intervals,
Low ($v$$>$-100 \kms) and High ($v$$<$-100 \kms) (LVI and HVI).
This analysis highlights many shock signatures such as \en\, maxima in the 
HVI of the
brightest knots and a sharp increase of \xe\, and \te\, in the HVI of the
strong knot L, which has a clear bow morphology in the HST images of 
\citet{reipurth97} 
and is the knot with the highest shock velocity in the jet beam 
\citep{hartigan01}.

We also analysed for the first time the faint jet powered by the young source 
Sz68. In this case we obtain low values of the electron
and total density (\en$\sim$200-10$^3$ \cmc\, and \nh$\sim$2-6 10$^3$ \cmc) 
and intermediate
values of the ionisation fraction and the temperature 
(\xe$\sim$0.05-0.25 and \te$\sim$0.8~10$^4$ K).
For this jet we also measured proper motions of $\sim$300-600 \kms\, 
from which, assuming
a jet radius of $\sim$100-300 AUs we derived a mass loss rate ranging between
a few 10$^{-8}$ and a few 10$^{-7}$\msolyr, i.e. of the same order of magnitude
 of \mjet\, found for typical HH jets \citep{hartigan94,bacciotti99,podio06}. 

Using the inferred physical parameters we derive an estimate for the 
dust content along the jets by comparing observed and predicted ratios between 
calcium and sulphur lines.
The derived Ca gas-phase abundance is lower than Solar 
([Ca]$_{gas}$/[Ca]$_{Solar}$$\sim$0.2-0.85) in all the jets, with the 
exception of HH 83 to which, however, no dereddening has been applied, thus
yielding upper limits to the Ca gas-phase abundance.
In general,
the measured Ca depletion is lower with respect to 
the interstellar medium ([Ca]$_{gas}$/[Ca]$_{Solar}$$\sim$0.01 in Orion, 
\citealt{baldwin91}).
This result suggests that mild shocks occurring along the jets 
are partially destroying dust grains, thus releasing calcium atoms into 
gaseous form.
Interestingly, a higher degree of depletion is estimated in the low-excited
jets (i.e. HH 111 and HH 1), while small or no depletion is found along
HH 83 and HH 24 which show higher ionization fractions and temperatures.
This seems to indicate that the efficiency of shocks in 
reprocessing dust grains strongly increases with shock 
velocity, as expected from theoretical models 
\citep[e.g., ][]{jones94,guillet09}.
This idea is also supported by the finding that along HH 111 the
depletion is minimum in the HVI of knot L, which has the largest shock velocity
\citep{hartigan01}. 

The measured values of Ca gas-phase abundance, however, are too large
to be explained by dust reprocessing in shocks which are slower than 
100 \kms\, \citep{jones94,may00,guillet09}.
This discrepancy between observational results and predictions of 
theoretical models may indicate that the material in the jet is 
extracted from a wide region of the disk that includes both the inner region
close to the star, inside the so-called dust evaporation radius, R$_{evp}$, 
where the dust is destroyed by the stellar radiation \citep{isella05},
and the region beyond R$_{evp}$ from where the dust is lifted and accelerated
in the flow.
Modeling, however, is required to test this idea.

\begin{acknowledgements}
      We are grateful to G. Pineau des For{\^e}ts, V. Guillet, and S. Cabrit for
      useful discussions on dust reprocessing in shocks.
      We are also grateful to the referee for the attentive comments that 
      allowed us to improve the first version of this paper.
      We thank the Irish Research Council for Science, Engineering and 
      Technology which funded the work of Linda Podio. This work was partially 
      supported by the European Community's Marie Curie Research and Training 
      Network JETSET (Jet Simulations, Experiments and Theory) under contract 
      MRTN-CT-2004-005592. 
      TPR was partially supported by Science Foundation Ireland through their
      Research Frontiers Programme.
\end{acknowledgements}

\Online

\begin{appendix} 
\section{Physical structure of previously analysed jets}

In this appendix we present the results obtained by 
applying spectral diagnostics to HH 1 (jet and bow), HH 2, HH 83, and HH 24.
These jets have been already analysed in previous papers with the same
diagnostics, but are re-analysed here with higher spatial sampling
($\Delta$x$\sim$1.2\arcsec) and an improved version of the diagnostic code 
as explained in Sect.~\ref{sect:diag}. 
The values presented are used for the investigation of the dust reprocessing 
in these jets (see Sect.~\ref{sect:dust}).

\subsection{HH 1/2: the jet and its terminal bows}
\label{sect:hh1}
   \begin{figure}
     \centering
     \includegraphics[width=8.cm]{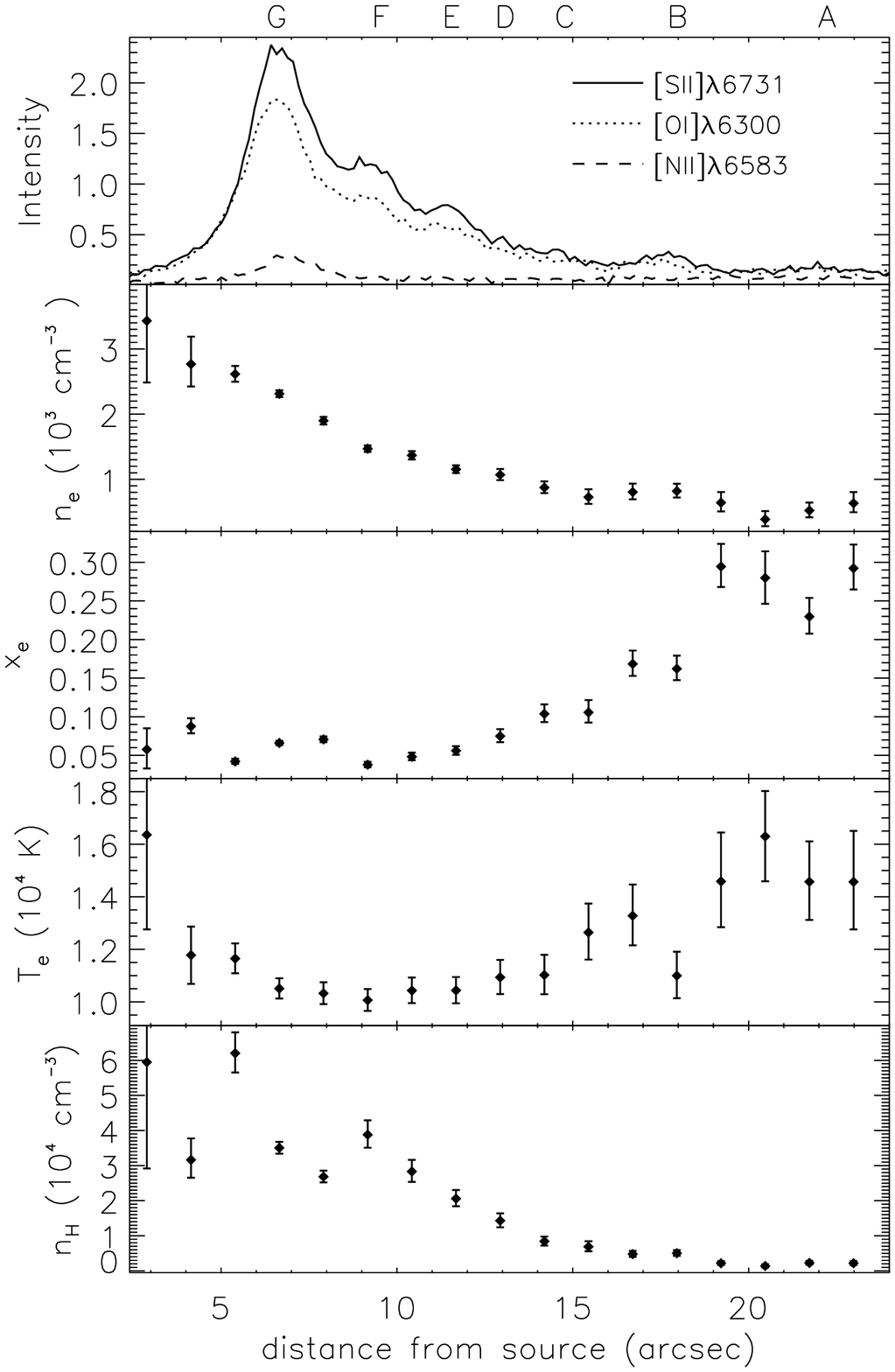}
   \caption{Variation of the physical parameters for the HH 1 jet beam as a 
function of the distance from the source.
{\em From top to bottom panel:} 
intensity profiles of the optical lines, the electron density, \en, in units
of 10$^3$ \cmc, the ionisation fraction, \xe, the temperature, \te, in units
of 10$^4$ K, and the total density, \nh, in units of 10$^4$ \cmc.}\label{hh1_phys}
    \end{figure}

   \begin{figure*}
     \centering
     \includegraphics[width=14.cm]{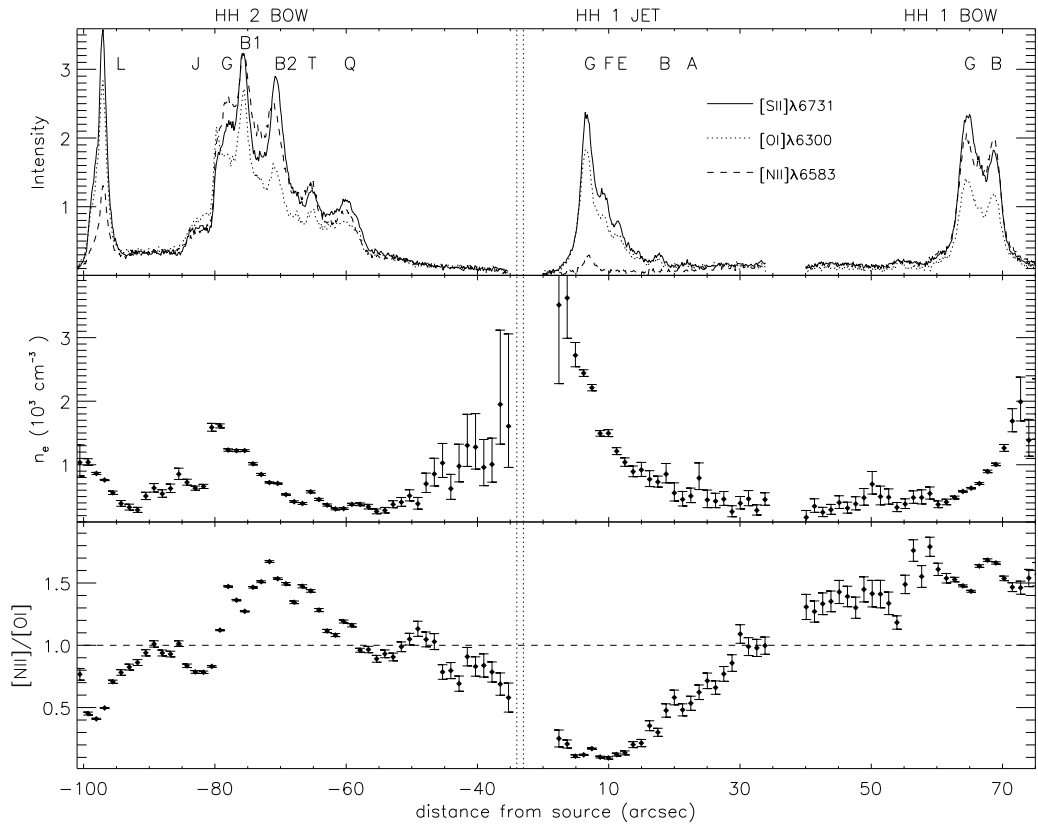}
   \caption{Analysis of the excitation conditions in the HH 1 jet and its
terminal bows HH 1 and HH 2 as a 
function of the distance from the source.
{\em From top to bottom panel:} 
intensity profiles of the optical lines, 
the electron density, \en, in units of 10$^3$ \cmc, 
and the \nii/\oi\, ratio.}\label{bow1_2_phys}
    \end{figure*}

%

The physical structure of the HH 1 jets and its terminal bows has been 
previously analysed by \citet{solf91}, \citet{nisini05}, and 
\citet{bohm85}, \citet{solf88}, \citet{eisloffel94}, \citet{bally02}.
In our observations the slit is aligned along the jet and it covers the fainter
western/eastern part of HH 1 and HH 2 respectively, 
i.e. the knots B and G, which are located to the west of the
HH 1 bow apex (knot F), and the knots L, J, G, B, T, and Q 
located to the east of the bright H and A knots in HH 2.

The results obtained by applying the BE technique to the HH 1 jet are shown in
Fig.~\ref{hh1_phys}. 
The values for the physical parameters are in a good agreement with the ones 
obtained in \citet{nisini05} but here the spatial sampling is two times better.

We could not apply the diagnostic to the terminal bows.
The BE technique, in fact, relies on the assumption of low excitation
conditions, i.e. S is all ionised but there is no S$^{++}$, and oxygen 
and nitrogen are ionised at most once \citep{bacciotti99}.
This assumption is satisfied when studying the jet beam where the gas interacts
with material that was already put into motion from the passage of previous
shocks, but may not be correct when dealing with the terminal bows.
This picture is confirmed by proper motions studies by \citet{bally02} which
indicate shock velocities lower than 30 \kms\, in the jet and velocity 
variations of up to 100-200 \kms\, in the bows.
Moreover,  \citet{bohm85} and \citet{solf88} detected many
high excitation lines in HH 1 and HH 2 such as O$^{++}$, S$^{++}$, and ArIV.

Even if in our case the slit is covering the lateral part of the bows
the excitation conditions may still be too high.
This is indicated by the detection of the high excitation \ariii\lam7135.8 
line in our spectra and by the fact that the \nii\, lines, 
which are faint in the jet beam, show strong emission in the bows, 
comparable to the \sii\, lines (see the upper panel of Fig.~\ref{bow1_2_phys}).

In order to compare the physical conditions in the jet and in the bows 
we computed the electron density, \en, 
and the \nii/\oi\, ratio along all the slit length.
The electron density does not show large differences.
It varies between 0.05-4~10$^3$~\cmc\, along all the jet.
While in the jet beam \en\, is decreasing with distance from the source, 
both in the HH 1 and in HH 2 bows, on the contrary, 
there is a decreasing trend going from the shock apex toward the source.
This is expected behind a shock front and confirms the results found by
\citet{bohm85} and \citet{solf88}.
Our values of \en\, are lower than those inferred by \citet{bohm85}, however,
because of the different slit alignment (on the brightest spots at the apex 
of HH 1 and HH 2 for \citealt{bohm85}, and on the bows wings in our case)
and show that the density in the two bows is maximum at the apex and is 
fading towards the wings.
The \nii/\oi\, ratio is a good indicator of the excitation conditions and, 
in particular, of the ionisation state of the gas.
Fig.~\ref{bow1_2_phys} shows that \nii/\oi\, is $<$1 in the jet beam
and in knot L, while it is $>$1 in the HH 1 and HH 2 bows, indicating
that the excitation is much higher in the bows, and a higher value of
\xe\, is expected.

\subsection{HH 83: the physical structure of the inner knots}
\label{sect:hh83}
   \begin{figure}
     \centering
     \includegraphics[width=8.cm]{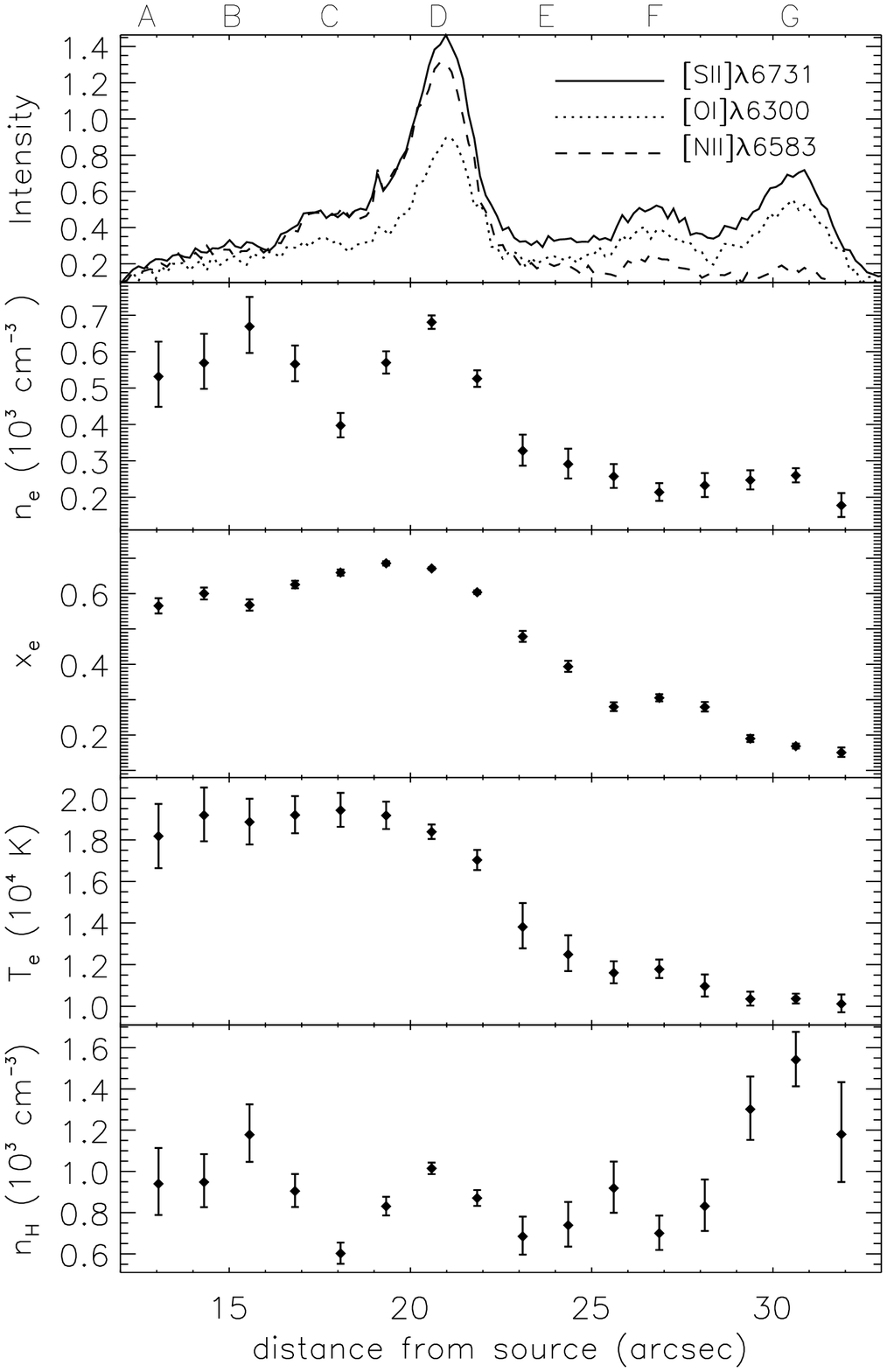}
   \caption{Variation of the physical parameters for the HH 83 jet as a 
function of the distance from the source.
{\em From top to bottom panel:} 
intensity profiles of the optical lines, the electron density, \en, in units
of 10$^3$ \cmc, the ionisation fraction, \xe, the temperature, \te, in units
of 10$^4$ K, and the total density, \nh, in units of 10$^3$ \cmc.}\label{hh83_phys}
    \end{figure}


The physical structure of the HH 83 jet has been already derived in 
\citet{podio06}.
Thanks to the high S/N of these data, however, we obtained a
sampling which is four times larger with respect to previous results.
Moreover the good quality of the data, which allowed us to properly 
subtract the continuum emission from the reflection nebula 
Re 17 \citep{rolph90},
and the use of the ``improved'' diagnostic code
allowed us to estimate the
gas physical conditions in the inner part of the jet, where the emission in the
\nii\, lines is comparable to the \sii\, and \oi\, emission.

The derived parameters indicate that the excitation conditions are very high
in the knots near to the source (knots from A to D) 
with \en$\sim$400-700 \cmc, and high values of the ionisation fraction
and temperature (\xe$\sim$0.4-0.7, and \te$\sim$1.5-2~10$^4$ K).

\subsection{The HH 24 jets}
\label{sect:hh24}
   \begin{figure*}
     \centering
     \includegraphics[width=14.cm]{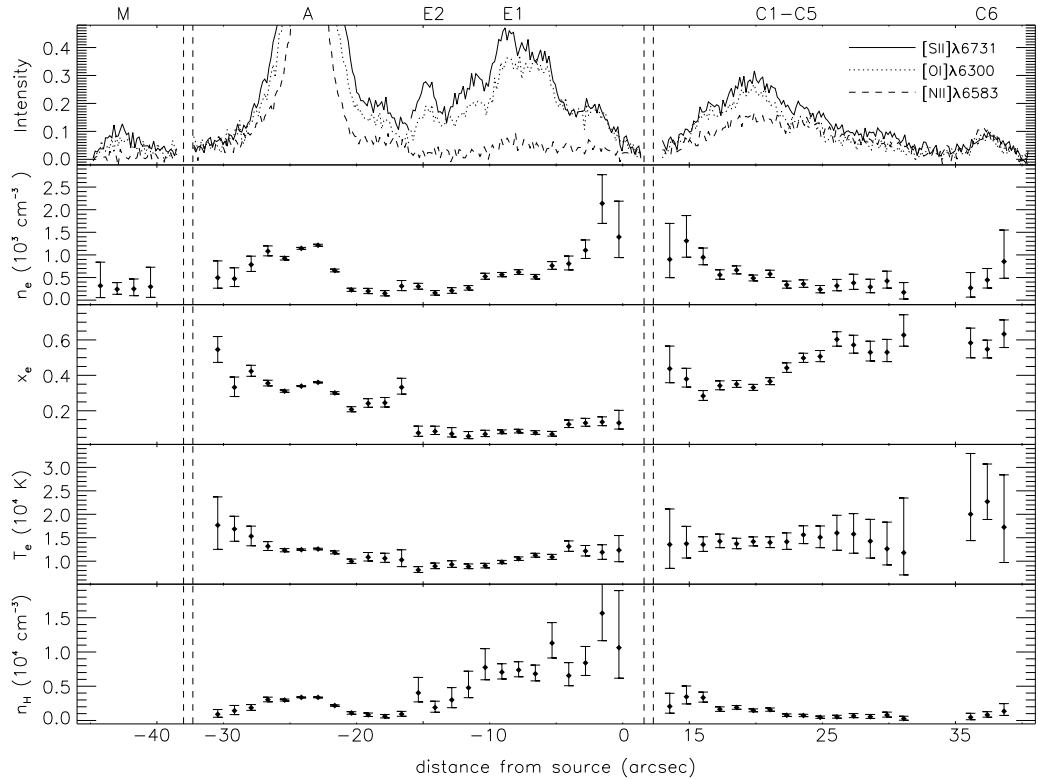}
   \caption{Variation of the physical parameters for the HH 24 C/E jet as a 
function of the distance from the source.
{\em From top to bottom panel:} 
intensity profiles of the optical lines, the electron density, \en, in units
of 10$^3$ \cmc, the ionisation fraction, \xe, the temperature, \te, in units
of 10$^4$ K, and the total density, \nh, in units of 10$^4$ \cmc.}\label{hh24_phys}
    \end{figure*}


The physical parameters along the jets HH 24 C, E, and A, has been already
derived in \citet{bacciotti99} and \citet{podio06}. 
In our observations the slit has been aligned along the axis HH 24 M-A-E and
thus only partially covers the knots of the HH 24 C jet up to 40$''$ (knot C6).
This is why the line profiles in the upper panel of 
Fig.~\ref{hh24_phys} show fainter emission in the knots of group C, 
contrary to what was found in \citet{podio06}, where the slit was aligned
along the HH 24 C jet.  
The variation of the physical parameters obtained by applying the BE technique
is shown in Fig.~\ref{hh24_phys}.
The sampling is improved of around one third with respect to previous analyses 
\citep{bacciotti99, podio06} 
allowing us to highlight the different excitation conditions in the
various groups of knots detected in the HH 24 complex 
(HH 24 A, HH 24 M/E, and HH 24 C) 
and supporting the idea that these knots may belong to different jets. 

\end{appendix}

\end{document}